# Exciton Spin Relaxation Time in Quantum Dots Measured by Continuous-Wave Photoluminescence Spectroscopy

S. Mackowski*, T. A. Nguyen, H. E. Jackson, and L. M. Smith

*Department of Physics, University of Cincinnati, 24551-0011 Cincinnati OH*

J. Kossut, and G. Karczewski

*Institute of Physics Polish Academy of Sciences, Warszawa, Poland*

*We demonstrate a new method of measuring the exciton spin relaxation time in semiconductor nanostructures by continuous-wave photoluminescence (PL). We find that for self-assembled CdTe quantum dots (QDs) the degree of circular polarization of emission is larger when exciting polarized excitons into the lower energy spin state ($\sigma^-$ – polarized) than in the case when the excitons are excited into the higher energy spin state ($\sigma^+$ – polarized). A simple rate equation model gives the exciton spin relaxation time in CdTe QDs equal to $\tau_S=4.8\pm0.3$ ns, significantly longer than the QD exciton recombination time $\tau_R=300$ ps.*

* corresponding author e-mail: seb@physics.uc.edu

In order to utilize the spin degree of freedom in real spintronics devices, a thorough understanding of the spin dynamics in semiconductor materials is required [1,2]. Semiconductor nanostructures, particularly self-assembled quantum dots (QDs), are of great interest in this regard because of the very long spin coherence times that could allow control and manipulation of the electron spin [3]. The robustness of spin in QDs is a direct consequence of the discrete density of electronic states resulting from the strong spatial confinement in all three dimensions [4]. This strong confinement reduces significantly (by several orders of magnitude) elastic spin dissipation processes related to spin-orbit interaction that are predominantly responsible for the very rapid spin relaxation in semiconductor quantum wells [5]. Recent time-resolved photoluminescence (PL) measurements performed on QD ensembles show that indeed the exciton spin relaxation time ($\tau_S$) in QDs can be longer than the exciton recombination time ($\tau_R$) [6]. Similar qualitative conclusions have also been drawn from the results of single dot PL spectroscopy [7-9]. However, estimating the actual value of $\tau_S$ from these time-resolved PL experiments has been difficult. Due to the significant difference between $\tau_R$ and $\tau_S$ in QDs, the degree of polarization shows essentially no measurable decay within the range of the exciton recombination time [6]. On the other hand, at times characteristic of spin relaxation in QDs the PL intensity of exciton recombination is usually weak so that the signal to noise ratio is insufficient to resolve accurately the value of $\tau_S$ [6].

A determination of the spin relaxation time can also be achieved by resonant excitation of the QDs ground states [6]. The resonant excitation reduces the number of possible spin relaxation mechanisms, such as coupling to the excited states. However, when the size of QDs is very small, the energy splitting between excited and ground states in QDs could be much larger than LO phonon energy. In this case one may be able to resonantly excite QD ground states through LO phonon-assisted absorption [10]. Such an excitation process would not lead to any significant contribution of excited state scattering to the exciton spin relaxation in QDs.

In this letter we present a method which allows a direct measurement of $\tau_S$ of the excitons in QDs by means of continuous-wave (CW) PL spectroscopy. We find that in an applied magnetic field, when the exciton spin levels in QDs are split, the degree of the polarization is larger when creating spin-polarized excitons in the lower energy level ($\sigma^-$ – polarized) than when creating spin-polarized excitons in the upper energy level ($\sigma^+$ – polarized). This difference, determined by the Boltzmann factor ($\Delta E/kT$, where $\Delta E$ is the splitting between spin states in QDs), increases with increasing magnetic field. A simple rate equation model gives the value of $\tau_S$ equal to 4.8±0.3 ns, an order of magnitude longer than $\tau_R$ in these QDs. We find that the result for the particular system is very sensitive to the relation between $\tau_S$ and $\tau_R$.

The sample containing CdTe QDs embedded in a ZnTe matrix was grown by molecular beam epitaxy on a (100) oriented GaAs substrate. The self-assembled CdTe QDs were formed by depositing 4 monolayers of CdTe on a ZnTe buffer layer at a temperature of 320C. Further details of the growth can be found elsewhere [7]. The PL spectrum of this QD sample displays a broad emission centered at an energy of 2.107 eV with a linewidth of 70 meV. The rather large width of this line originates from the distribution of sizes and/or alloy fluctuations among QDs in the ensemble.

In order to study the spin relaxation of excitons confined to the CdTe QDs we used CW PL spectroscopy in a magnetic field. The sample was mounted to the cold finger in a continuous flow helium cryostat at T=5 K. Magnetic fields up to 4 T were applied in a Faraday configuration. We used an Ar ion pumped dye laser (Rhodamine 6G) to resonantly excite the QD ground states. The excitation was selected to be $\sigma^+$ or $\sigma^-$ – polarized by combining a Glan-Thompson linear polarizer with a Babinet-Soleil compensator. A similar configuration was used to analyze the polarization of the QD emission. The signal was dispersed through a DILOR triple spectrometer working in a subtractive mode and detected by a liquid nitrogen cooled CCD detector.

In Fig. 1a we show resonantly excited CW-PL spectra of CdTe QDs obtained at B=3 T with $\sigma^+$ – polarized excitation. Emission in both $\sigma^+$ and $\sigma^-$ circular polarizations is shown. The three broad spectral features observed for each spectrum originate from QDs which have been populated by spin-polarized excitons through a LO phonon-assisted absorption directly into their ground states [10]. When the excitation is $\sigma^-$ – polarized ($\sigma^+$ – polarized) the QD emission is also predominantly $\sigma^-$ – polarized ($\sigma^+$ – polarized). As shown by these measurements, in a magnetic field the circular polarization of the QD emission is predominantly the same as that of the excitation.

In order to evaluate the exciton spin relaxation time in CdTe QDs, we define the polarization, $P=(I^+-I^-)/(I^++I^-)$, where $I^+$ and $I^-$ represent intensities of $\sigma^+$ and $\sigma^-$ – polarized emission, respectively. The energy dependence of the polarization for both circularly polarized excitations can be obtained directly from PL spectra similar to those shown in Fig. 1a. In Fig. 1b we show the polarization, P, for both $\sigma^+$ and $\sigma^-$ – polarized excitations measured at three magnetic fields: 0, 1 and 4 T. At B=0 T, as expected, the PL polarization is zero at all photon energies, regardless of the polarization of the excitation. However, with increasing magnetic field, the magnitude of the polarization of all three LO phonon replicas increases monotonically reaching approximately 60% at 4T. While it appears that the polarization P decreases with the number of LO phonons emitted, this is an artifact resulting from the unpolarized broad emission associated with direct excited state-ground state excitations. A detailed analysis including the results of single QD PL spectroscopy has shown that the polarization is essentially identical for all LO phonon replicas [11].

The spectra presented in Fig. 1b indicate that the spin dynamics of excitons in QDs depends significantly on the degeneracy of the exciton spin levels. When the exciton spin levels are degenerate (B=0 T) the spin-polarized excitons created resonantly in QDs randomize their spins very efficiently before they recombine ($\tau_S$ is much shorter than $\tau_R$). In this case the observed polarization is zero for both $\sigma^+$ and $\sigma^-$ – polarized excitations. In contrast, when an external magnetic field is applied, the degeneracy of exciton spin levels is lifted, and $\tau_S$ becomes much longer than $\tau_R$. Therefore, for non-degenerate exciton spin levels in QDs, the PL polarization is significant *only* for QDs excited through LO phonon-assisted absorption.

However, as seen in Fig. 1b, the magnitude of P for all three LO phonon replicas is clearly larger for $\sigma^-$ – polarized excitation than for $\sigma^+$ – polarized excitation. We find that this difference, $\Delta P = |P^-| - |P^+|$, increases with magnetic field and appears to saturate at a magnetic field of approximately 4 T. We note that by analyzing the difference $\Delta P$ (instead of the polarization itself), we restrict our discussion *only* to QDs populated through LO phonon-assisted absorption. This observation of a monotonically increasing $\Delta P$ allows a sensitive measurement of the exciton spin relaxation time in CdTe QDs.

From the above results it is clear that the spin relaxation time from the lower spin state ($\sigma^-$ – polarized) to the upper spin state ($\sigma^+$ – polarized) is *not* the same as for scattering from the upper spin state to the lower one. Indeed this asymmetry is clearly due to the partial thermalization of excitons to the lower energy state. We use a simple two-level rate equation model (see Fig. 2) to analyze the experimental results shown in Fig. 1b:

$$\frac{dn_+}{dt} = G^+ - \frac{n_+}{\tau_R} - \frac{n_+}{\tau_S} + n_- \frac{e^{-\Delta E/kT}}{\tau_S}$$

$$\frac{dn_-}{dt} = G^- - \frac{n_-}{\tau_R} + \frac{n_+}{\tau_S} - n_- \frac{e^{-\Delta E/kT}}{\tau_S}$$

Here $n_+$ and $n_-$ represent the occupation of the upper ($\sigma^+$ – polarized) and lower ($\sigma^-$ – polarized) energy levels. $G^+$ ($G^-$) is the generation rate of $\sigma^+$ ($\sigma^-$) – polarized excitation. $\Delta E = g_{eff} \mu_B B$ is the magnetic field-induced Zeeman splitting between the exciton spin levels, where $g_{eff}$ is the effective exciton g-factor in the QD. Previous single-dot PL measurements on this sample have shown that the $g_{eff}$ of these QDs is $-3$ [12] and is very uniform over the whole QD ensemble (for QDs with different energies). The exciton recombination time, $\tau_R$, has also been measured independently and is equal to 300 ps for these CdTe QDs [7]. For simplicity, we assume that for a given polarization of the excitation, only a single exciton spin state is populated. In other words, for $\sigma^+$ – polarized excitation the value of $G^+ = 1$ and $G^- = 0$, while for $\sigma^-$ – polarized excitation $G^+=0$ and $G^- =1$. Solving these equations for steady state, we find that solutions for $\sigma^+$ – polarized excitation take the form:

$$n_+^+ = \frac{\tau_R(\tau_R + \tau_S e^{\Delta E/kT})}{\tau_R + e^{\Delta E/kT}(\tau_R + \tau_S)} \quad \text{and} \quad n_-^+ = \frac{\tau_R^2 e^{\Delta E/kT}}{\tau_R + e^{\Delta E/kT}(\tau_R + \tau_S)}.$$

The upper index is the polarization of the excitation. On the other hand, for $\sigma^-$ – polarized excitation we obtain:

$$n_+^- = \frac{\tau_R^2}{\tau_R + e^{\Delta E/kT}(\tau_R + \tau_S)} \quad \text{and} \quad n_-^- = \frac{\tau_R e^{\Delta E/kT}(\tau_R + \tau_S)}{\tau_R + e^{\Delta E/kT}(\tau_R + \tau_S)}.$$

A little algebra shows that difference between P⁺ and P⁻ takes a particularly simple form:

$$\Delta P = |P^-| - |P^+| = \frac{n_-^- - n_+^-}{n_-^- + n_+^-} - \frac{n_+^+ - n_-^+}{n_+^+ + n_-^+} = \frac{2\tau_R (e^{\Delta E/kT} - 1)}{\tau_R + e^{\Delta E/kT}(\tau_R + \tau_S)}.$$

Knowing the Zeeman splitting ΔE of the excitons in the CdTe QDs [12], we can easily fit this expression to our experimental data. The solid line in Fig. 3 shows the result of the fit, while the points are the values obtained experimentally for *all three* LO phonon replicas seen in the PL spectra. From the fit we obtain the value of the exciton spin relaxation time in CdTe QDs equal to 4.8±0.3 ns, that is more than order of magnitude longer than the exciton recombination time in these QDs. In addition the dependencies calculated for two other values of $\tau_S$ are shown (dashed and dotted lines). Note that this method results in a very sensitive measurement of $\tau_S$ even though $\tau_S$ is an order of magnitude longer than the recombination time of excitons confined to these QDs.

In conclusion, we have demonstrated a new method for determining the spin relaxation time of excitons in CdTe QDs by continuous-wave PL measurements. We observe a larger degree of circular polarization when excitons are resonantly excited into the lower energy spin state in QDs. This indicates that cooling of exciton spins is more effective than warming by a Boltzmann factor. Using a simple rate equation model, we find the spin relaxation time in CdTe QDs to be equal to 4.8±0.3 ns. The method described here can also be used for other systems where the spin relaxation time is longer than the exciton recombination time.

Acknowledgements

The work was supported by NSF grants nr 9975655 and 0071797 (United States). Partial support through grant PBZ-KBN-044/P03/2001 and project SPINOSA (Poland) is acknowledged.

FIG. 1. (a) Resonantly excited CW PL spectra of CdTe QDs at B=3 T. The excitation was $\sigma^+$ – polarized while the emission of both $\sigma^+$ and $\sigma^-$ polarization was measured. (b) Polarization of QD emission obtained at B=0, 1 and 4 T.

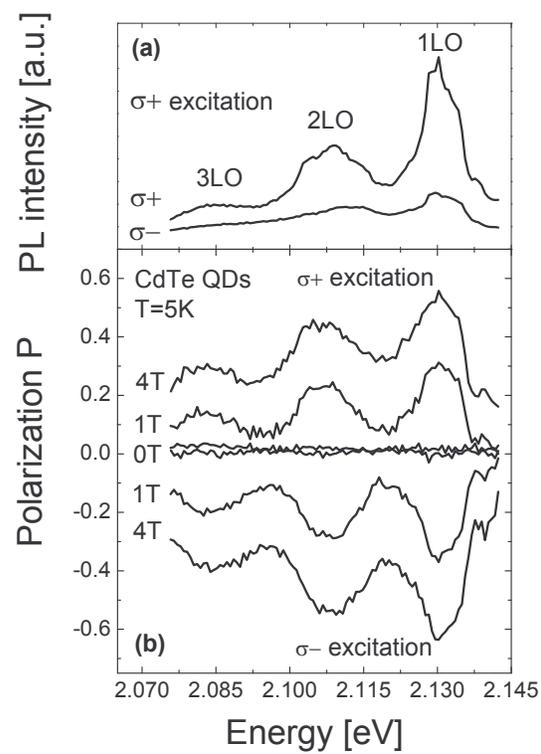

FIG. 2. Description of the symbols used in the rate equation model discussed in the text.

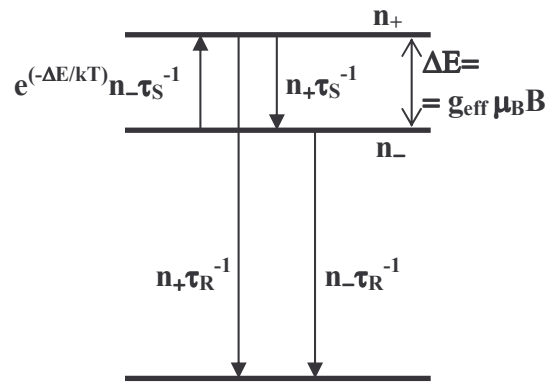

FIG. 3. Difference ΔP between polarizations measured for $\sigma^-$ and $\sigma^+$ - polarized excitations plotted as a function of ΔE/kT. The solid points represent experimental results for three LO phonon replicas. The lines are the result of fitting (solid line) and calculation for two different values of exciton spin relaxation time, $\tau_S$.

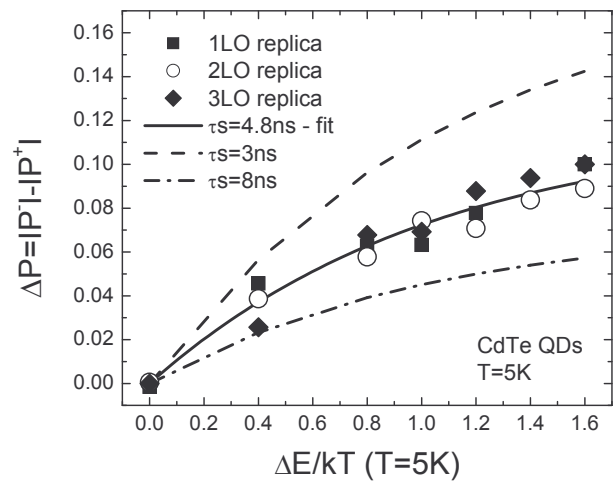